\documentclass[11pt,english,letterpaper,article,superscriptaddress,nofootinbib]{revtex4}

\usepackage{amsfonts}
\usepackage{graphicx}
\usepackage{float}
\usepackage{gensymb}

\newcommand{\pa} [1] {\sigma_{#1}}

\begin{document}
\title{Different Strategies for Optimization Using the Quantum Adiabatic Algorithm}

\author{Elizabeth Crosson}
\affiliation{Center for Theoretical Physics, Massachusetts Institute of Technology, Cambridge, MA 02139}
\affiliation{Department of Physics, University of Washington, Seattle, WA 98195}

\author{Edward Farhi}
\affiliation{Center for Theoretical Physics, Massachusetts Institute of Technology, Cambridge, MA 02139}

\author{Cedric Yen-Yu Lin}
\affiliation{Center for Theoretical Physics, Massachusetts Institute of Technology, Cambridge, MA 02139}

\author{Han-Hsuan Lin}
\affiliation{Center for Theoretical Physics, Massachusetts Institute of Technology, Cambridge, MA 02139}

\author{Peter Shor}
\affiliation{Center for Theoretical Physics, Massachusetts Institute of Technology, Cambridge, MA 02139}
\affiliation{Department of Mathematics, Massachusetts Institute of Technology, Cambridge, MA 02139}

\begin{abstract}
We present the results of a numerical study, with 20 qubits, of the performance of the Quantum Adiabatic Algorithm on randomly generated instances of MAX 2-SAT with a unique assignment that maximizes the number of satisfied clauses.  The probability of obtaining this assignment at the end of the quantum evolution measures the success of the algorithm.  Here we report three strategies which consistently increase the success probability for the hardest instances in our ensemble: decreasing the overall evolution time, initializing the system in excited states, and adding a random local Hamiltonian to the middle of the evolution.  
\end{abstract}

\maketitle

\section{Introduction}
The Quantum Adiabatic Algorithm (QAA) can be used on a quantum computer as an optimization method \cite{farhi-2000} for finding the global minimum of a classical cost function $f:\{0,1\}^n\rightarrow\mathbb{R}$.   The cost function is encoded in a problem Hamiltonian $H_P$ which acts on the Hilbert space of $n$ spin-$\frac{1}{2}$ particles,
\begin{equation}
H_P=\sum_{z\in\{0,1\}^n} f(z)|z\rangle\langle z|.\label{eq:Hp}
\end{equation}
The Hamiltonian $H_P$  is diagonal in the computational basis, and its ground state corresponds to the bit string that minimizes $f$.  To reach the ground state of $H_P$ the system is first initialized to be in the ground state of a beginning Hamiltonian, which is traditionally taken to be 
\begin{equation}
H_B=\sum_{i=1}^{n}\left(\frac{1-\sigma_{x}^{i}}{2}\right).  \label{eq:HB}
\end{equation}
The ground state of $H_B$, which can be prepared efficiently, is the uniform superposition of computational basis states  
\begin{equation}
|\psi_{init}\rangle=\frac{1}{\sqrt{2^n}} \sum_{z\in\{0,1\}^n} |z\rangle. \label{psi:init}
\end{equation}
The system is then acted upon by the time-dependent Hamiltonian
\begin{equation}
H(t)=(1-\frac{t}{T})H_B + \frac{t}{T}  H_P \label{eq:H}
\end{equation}
from time $t = 0$ to $t = T$ according to the Schrodinger equation 
\begin{equation}
i \frac{\mathrm{d}}{\mathrm{dt}} |\psi(t)\rangle\ = H(t) |\psi(t) \rangle. \label{eq:Schrodinger}
 \end{equation}
 For a problem instance with a unique string $w$ that minimizes $f$, the probability of obtaining $w$ at time $t = T$,
\begin{equation}P(T) = |\langle w | \psi(t = T) \rangle|^2 , \end{equation}
 is a metric for the success of the method on that particular instance.  

By the adiabatic theorem, if we prepare the system initially in the ground state of $H_B$ and evolve for a sufficiently long time $T$, then the state of the system at the end of the evolution will have a large overlap with the ground state of $H_P$.  Specifically, the adiabatic approximation requires $T > \mathcal{O}(g_{min}^{-2})$, where $g_{min}$ is the minimum difference between the ground state energy and the first excited state energy during the course of the evolution.     

In this paper we explore strategies that do not necessarily require a run time $T > \mathcal{O}(g_{min}^{-2})$.  We sidestep the usual question of determining how the run time $T$ needed to achieve a certain success probability scales with the input size $n$.  Instead we work at a fixed bit number, $n = 20$, and look at strategies for improving the success probability for hard instances at this number of bits.  We observe three strategies that increase the success probability for all of the hard instances we generated: evolving the system more rapidly (``{the hare beats the tortoise''}), initializing the system to be in a superposition of the states in the first excited subspace of $H_B$ (``{going lower by aiming higher}''), and adding random local Hamiltonian terms to the middle of the evolution path (``{the meandering path may be faster}''). 

\section{Instance Selection}
We sought to accumulate an ensemble of instances of MAX 2-SAT on $n = 20$ bits that are hard for the QAA as described above.  Our instances are constructed by randomly generating $60$ distinct clauses, each involving two distinct bits, and retaining the instance only if there is a unique assignment $w$ that minimizes the number of violated clauses.  We keep only those instances that have a unique minimal assignment because degenerate ground states of $H_P$ make the energy gap zero, and because we wish to avoid the complication of having success probabilities depend on the number of optimal solutions.   

We generated $202,078$ instances and selected all those having a low success probability at $T=100$, using $P(100) < 10^{-4}$ as our cutoff, resulting in a collection of 137 hard instances.  To speed up the search for these instances, we used a mean-field algorithm to approximate the QAA in equations \ref{eq:Hp} through \ref{eq:H} with $T = 100$, and then we discarded the instances that had a final mean-field energy of $0.5$ or less above the energy of the optimal assignment.   We checked that instances that are easy for the mean-field algorithm would also have a high success probability under the full Schrodinger evolution by sampling a separate population of $15,000$ instances, and found that whenever the mean-field algorithm produced a final energy less than $0.5$ above the ground state energy the instance had success probability $P(100) > 0.2$ according to the Schrodinger evolution.   The use of this filter allowed us to discard $3/4$ of the initial $202,078$ instances, and for the remainder we numerically integrated the Schrodinger evolution with $T = 100$.

\begin{figure}[H]
\begin{centering}
\includegraphics[scale=1.1,trim=0 5 0 0]{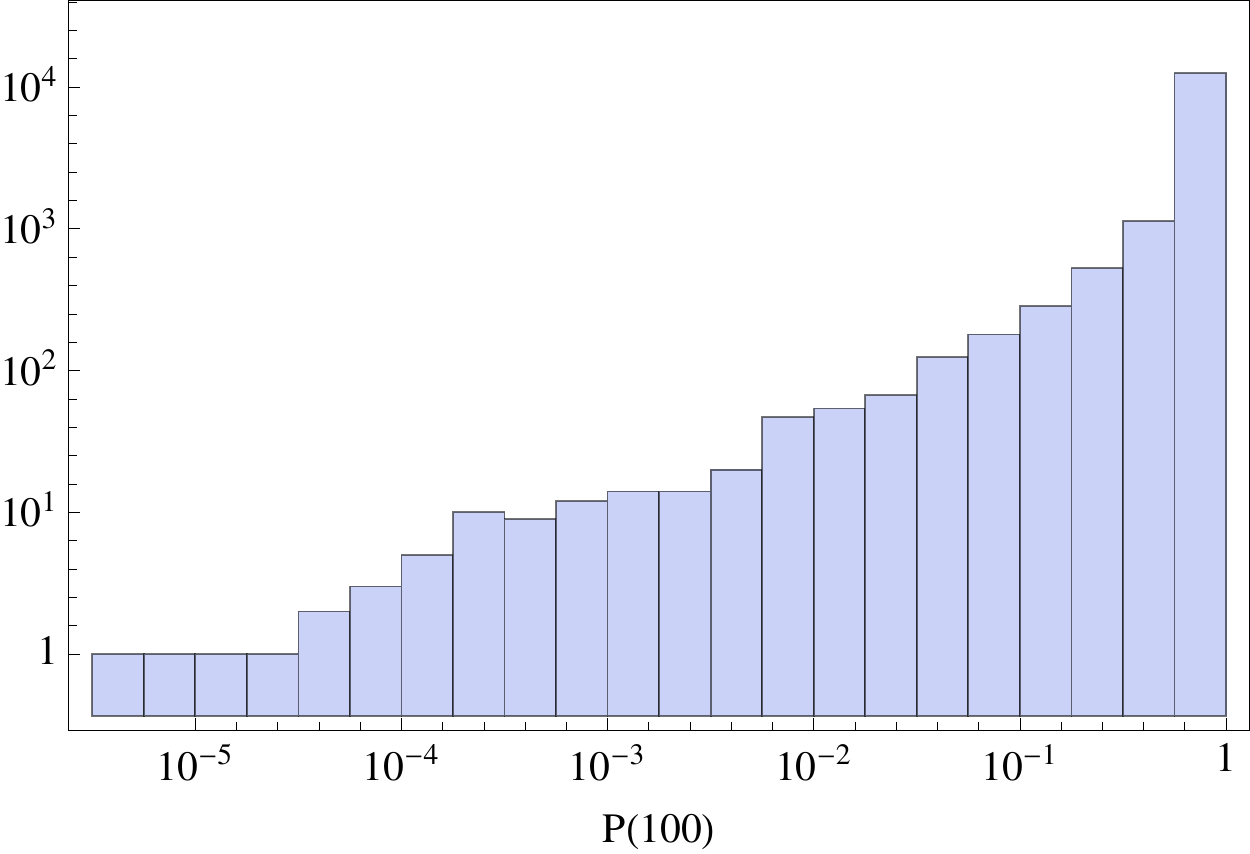}
\par\end{centering}
\caption{The distribution of success probabilities for 15000 instances. \label{fig:successDistribution}}
\end{figure}

The success probabilities at $T = 100$ for the test population of $15,000$ instances are given in figure \ref{fig:successDistribution}.   Most of the instances we generate have high success probability at $T = 100$ (in fact, over half of this population had $P(100) > 0.95$), and hard instances at this time scale and number of bits are rare.   This is why we needed to generate roughly $200,000$ total instances, and search through them using over 20,000 hours of CPU time, to obtain our ensemble of 137 instances which have $P(100) < 10^{-4}$. 
   
\section{The Hare Beats the Tortoise} 
\begin{figure}[h]
\begin{centering}
\includegraphics[scale=1.1,trim=20 10 0 0]{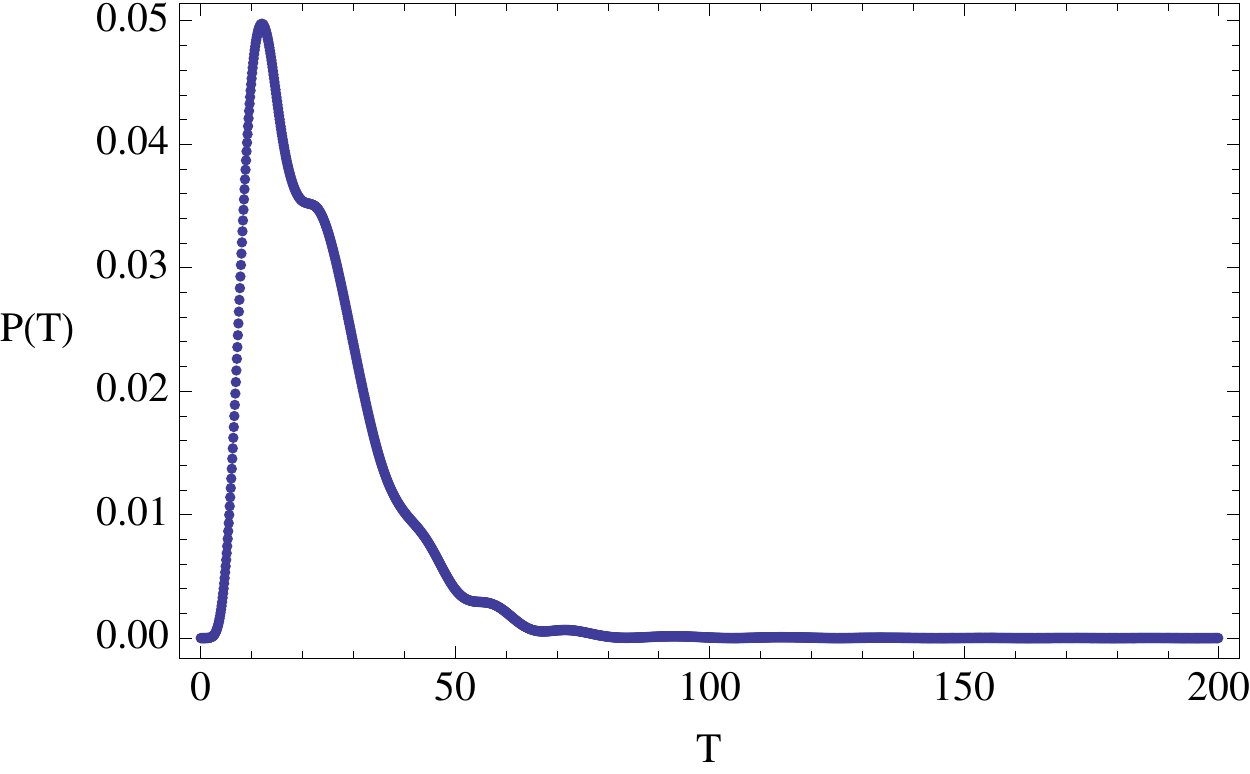}
\par\end{centering}
\caption{The success probability as a function of total evolution time $T$ for instance $\#1$.  \label{fig:successVSTotalT}}
\end{figure}

\vspace{-10pt}
Considering the success probability $P(T)$ as a function of the total evolution time $T$, we find that all of our instances with low success probability at $T = 100$ exhibit higher success probability at lower values of $T$.  Figure \ref{fig:successVSTotalT} depicts this phenomenon for a single hard instance, which happened to be the first instance we carefully examined. We will refer to this instance as instance $\#1$. We see a distinct peak of success probability  at $T_{\max} = 12$ with $P(T_{max})=0.05$, which is to be compared with $P(100) = 5\times10^{-5}$ and $P(200) = 5\times10^{-6}$.

\begin{figure}[h]
\begin{centering}
\includegraphics[scale=1.1,trim=38 10 0 1]{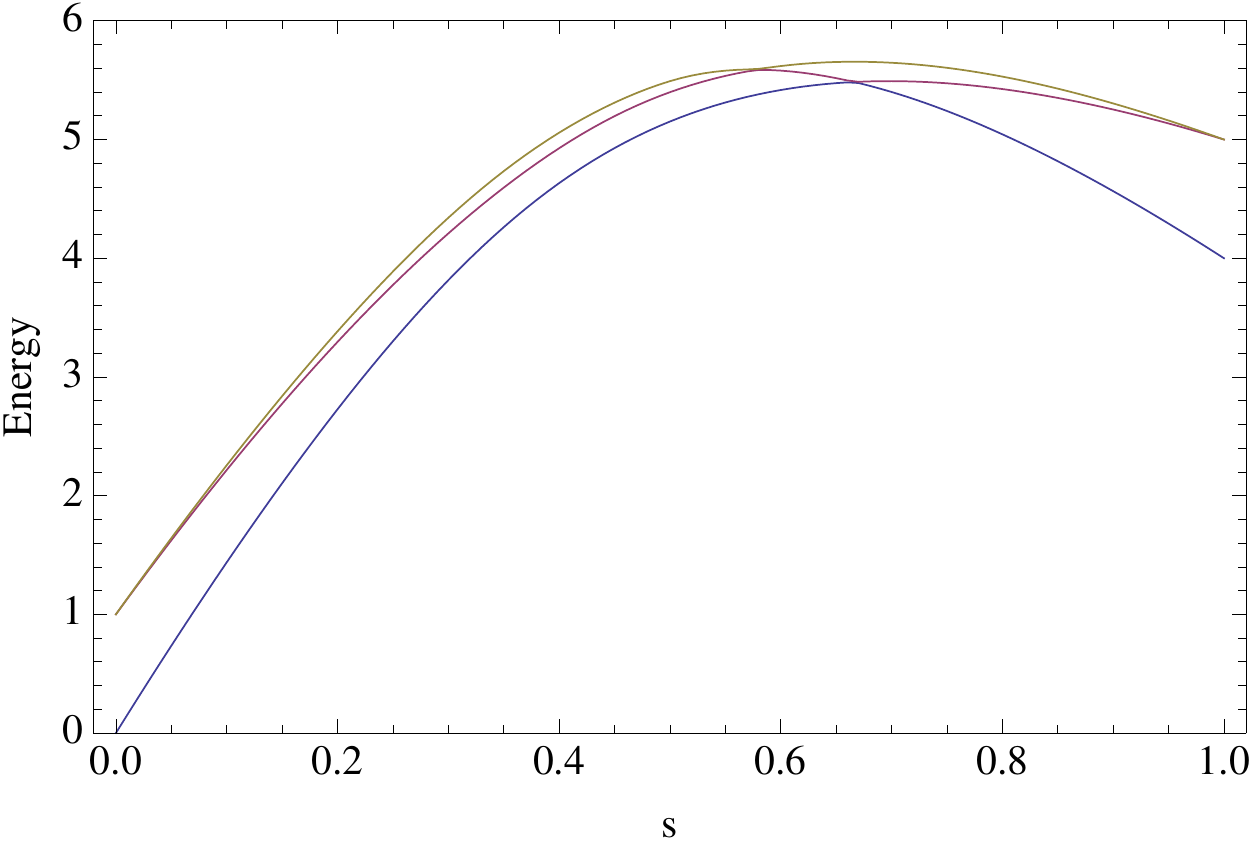}
\par\end{centering}
\caption{The lowest three energy levels for instance $\#1$.    \label{fig:spectrum137}}
\end{figure}
\begin{figure}[H]
\begin{centering}
\includegraphics[scale=1.55,trim=30 5 0 0]{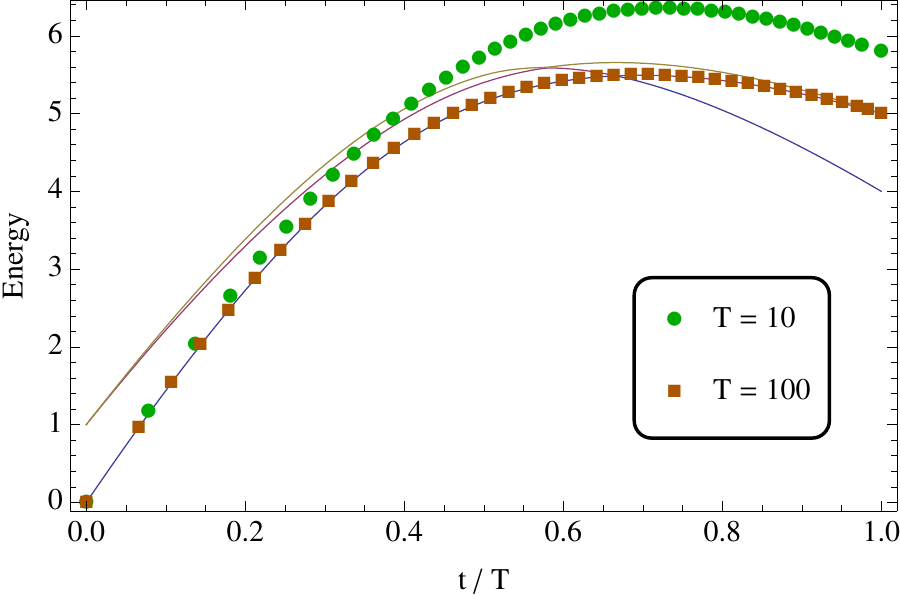}
\par\end{centering}
\caption{The lowest three energy levels for instance $\#1$, superimposed with the instantaneous expectation of the energy as a function of $t$ for $T = 10$ and $T = 100$.    \label{fig:energyExpectations}}
\end{figure}

In figure \ref{fig:spectrum137} we plot the three lowest energy levels of instance $\#1$, and we see a small energy gap which corresponds to an avoided crossing near $s = 0.66$.  To see why changing the Hamiltonian more rapidly increases the success probability, figure \ref{fig:energyExpectations} gives the instantaneous expectation of the energy, $\langle\psi(t)|H(t)|\psi(t)\rangle$, as a function of $t$ for $T = 10$ and $T = 100$, together with the three lowest energy eigenvalues.  We see that when the Hamiltonian is changed slowly, the $T = 100$ case, the system remains close to the ground state for all time $t < 0.66 T$, but then switches to closely following the first excited state after the avoided crossing, and arrives with most of its amplitude in the first excited state subspace of $H_P$ with virtually no overlap with $|w\rangle$. 

\begin{figure}[H]
\begin{centering}
\includegraphics[scale=1.6,trim=40 5 20 0]{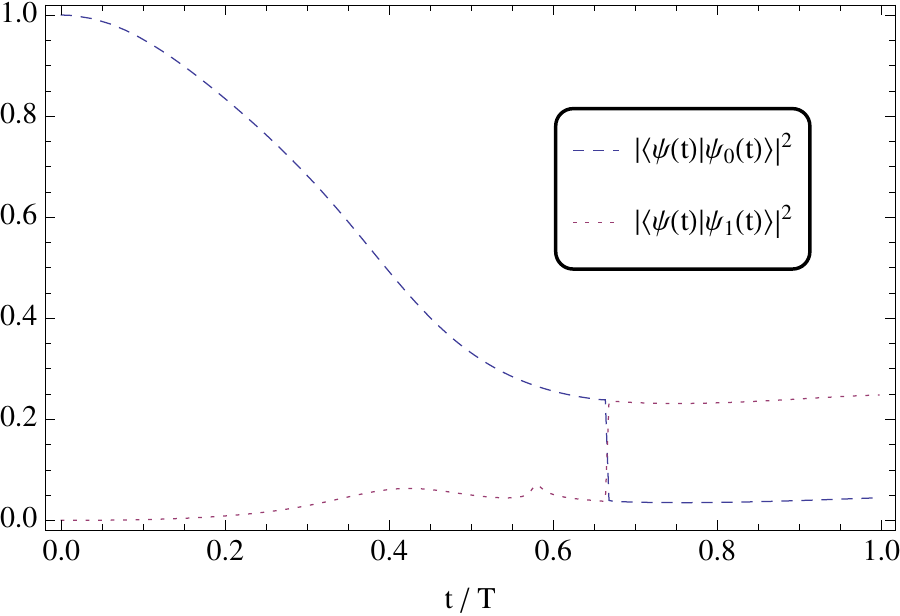}
\par\end{centering}
\caption{The overlap of the rapidly evolved system ($T = 10$) with the lowest two instantaneous energy eigenstates of $H(t)$, labeled here as $|\psi_0(t)\rangle$ and $|\psi_1(t)\rangle$.  The bump in the overlap with the first excited state near $s = 0.58$ coincides with the avoided crossing between levels 2 and 3, as seen in figure \ref{fig:spectrum137}.\label{fig:EnT}}
\end{figure}

In figure \ref{fig:EnT} we track the overlap of the rapidly evolved system ($T = 10$) with the lowest two energy eigenstates of $H(t)$, and see that the overlap of the system with the ground state immediately after the crossing corresponds to the overlap with the first excited state immediately before it.  When evolving more rapidly, leaking substantial amplitude into the first excited state prior to the crossing is responsible for the increased probability of finding the system in the ground state at the end of the evolution.
\vspace{10pt}

\begin{figure}[H]
\begin{centering}
\includegraphics[scale=1.1,trim=60 5 0 0]{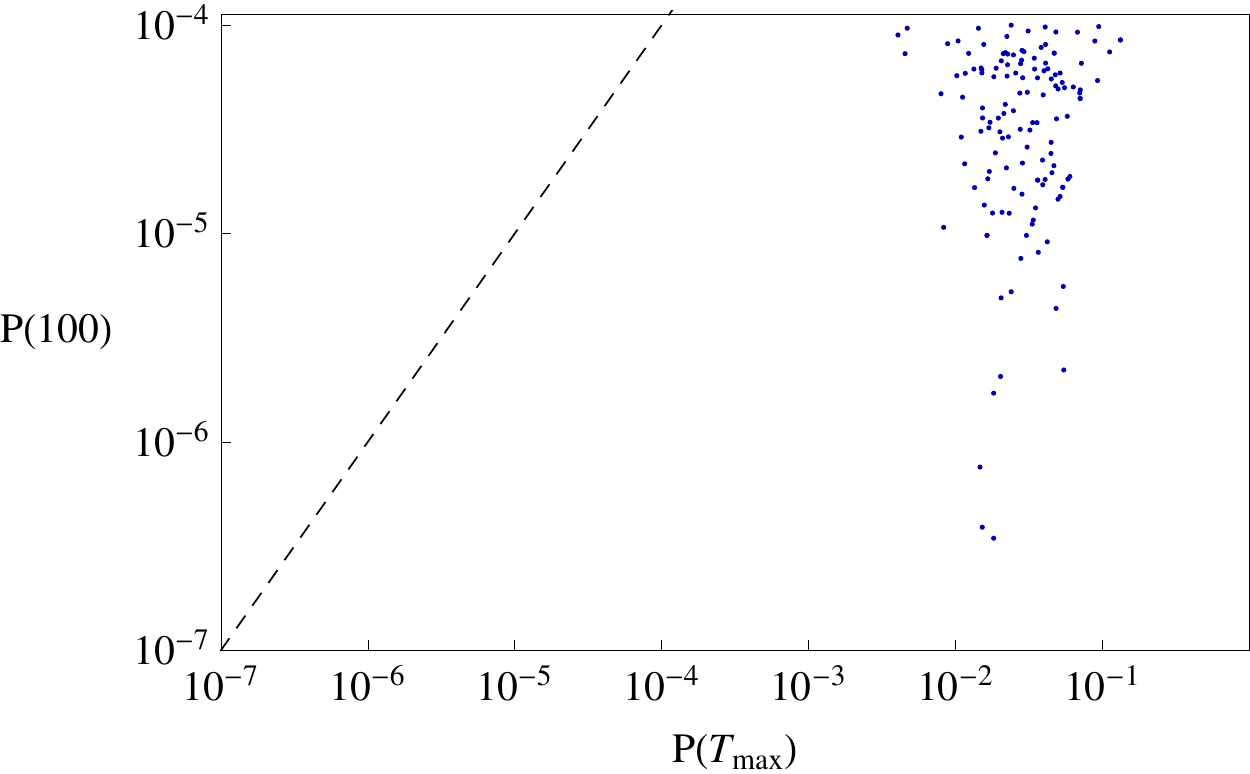}
\par\end{centering}
\caption{A log-log scatter plot comparing $P(T_{\max})$ with $P(100)$, where the value of $T_{\max}$ depends on the instance. \label{fig:shortTimesScatter}}
\end{figure}

Having described this phenomenon for a single instance, we now present evidence that it generalizes to many other hard instances.  For each of our 137 hard instances we determined the location $T_{\max}$ where the success probability is maximized in the interval $[0,40]$, and in figure \ref{fig:shortTimesScatter} we compare the success probability at $T_{\max}$ with the success probability at $T = 100$.  It is notable that every data point appears to the right of the $45\degree$ line, indicating that every one of our instances was improved by evolving the Hamiltonian more rapidly.  The minimum improvement $P(T_{\max})/P(100)$ for this batch of instances is $108$, and the median improvement is $809$.  

\begin{figure}[H]
\begin{centering}
\includegraphics[scale=1,trim=0 10 0 0]{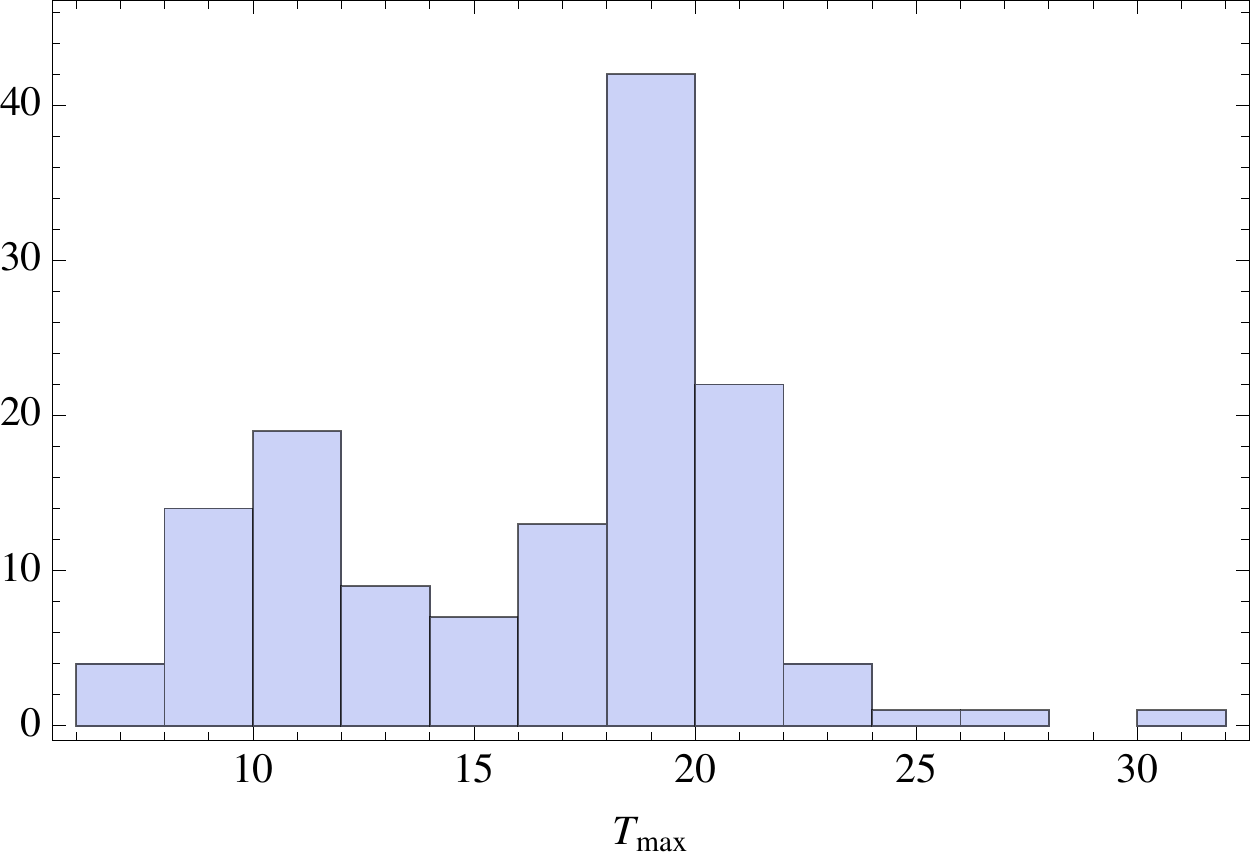}
\par\end{centering}
\caption{The distribution of the times $T_{max}$ at which the success probabilities of our hard instances are maximized in the interval $[0,40]$.\label{fig:histogramTmax}}
\end{figure}

From an algorithmic perspective, it may not be possible to efficiently estimate the value of $T_{\max}$ for each instance in advance.  The distribution of $T_{\max}$ for our 137 instances is shown in figure \ref{fig:histogramTmax}.  

\begin{figure}[H]
\begin{centering}
\includegraphics[scale=1.1,trim=60 5 0 0]{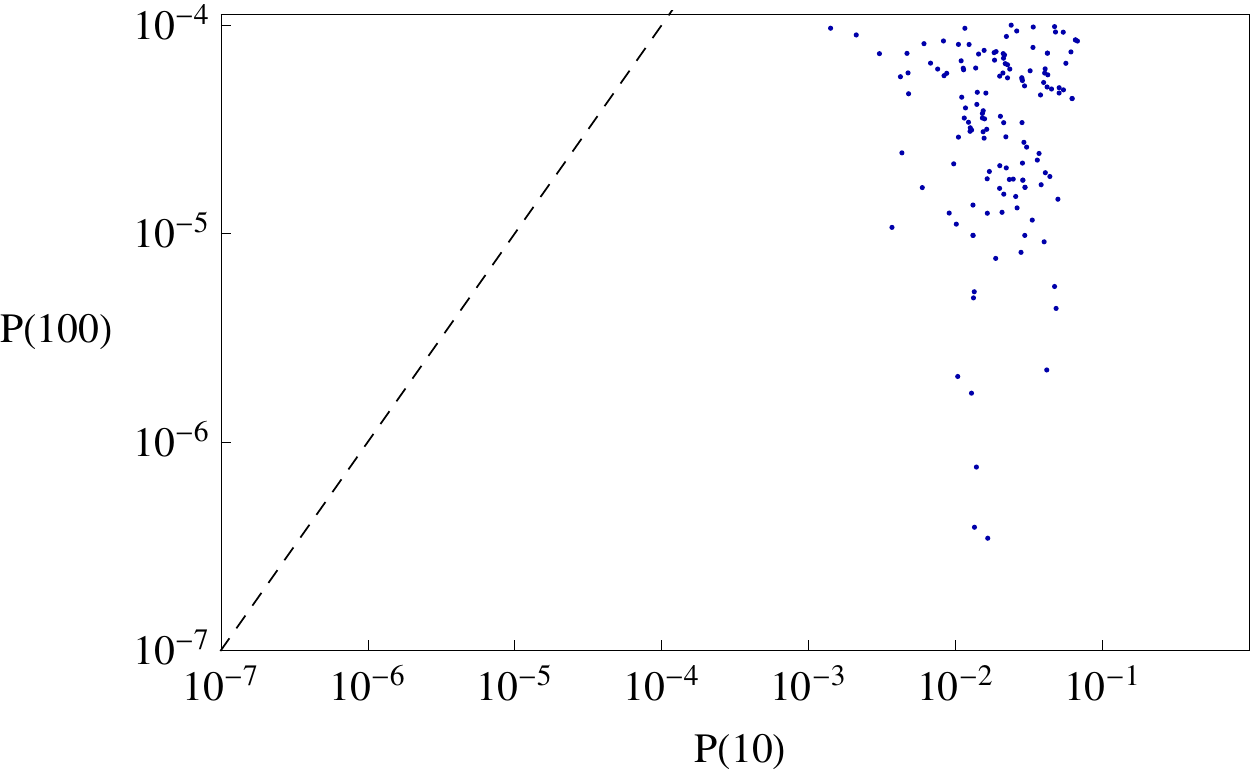}
\par\end{centering}
\caption{A log-log scatter plot comparing $P(10)$ with $P(100)$.  \label{fig:shortTimesScatter10}}
\end{figure}

\noindent The phenomenon we are describing is sufficiently robust that we can choose a fixed short time such as $T = 10$ and still gain a substantial improvement for every instance.  In figure \ref{fig:shortTimesScatter10} we compare the success probabilities at $T=10$ and $T=100$.  Here the minimum improvement $P(10)/P(100)$ is $15$, and the median improvement is $574$. 

\section{Going Lower by Aiming Higher \label{lowerAimingHigh}}
\begin{figure}[H]
\begin{centering}
\includegraphics[scale=1,trim=0 0 0 0]{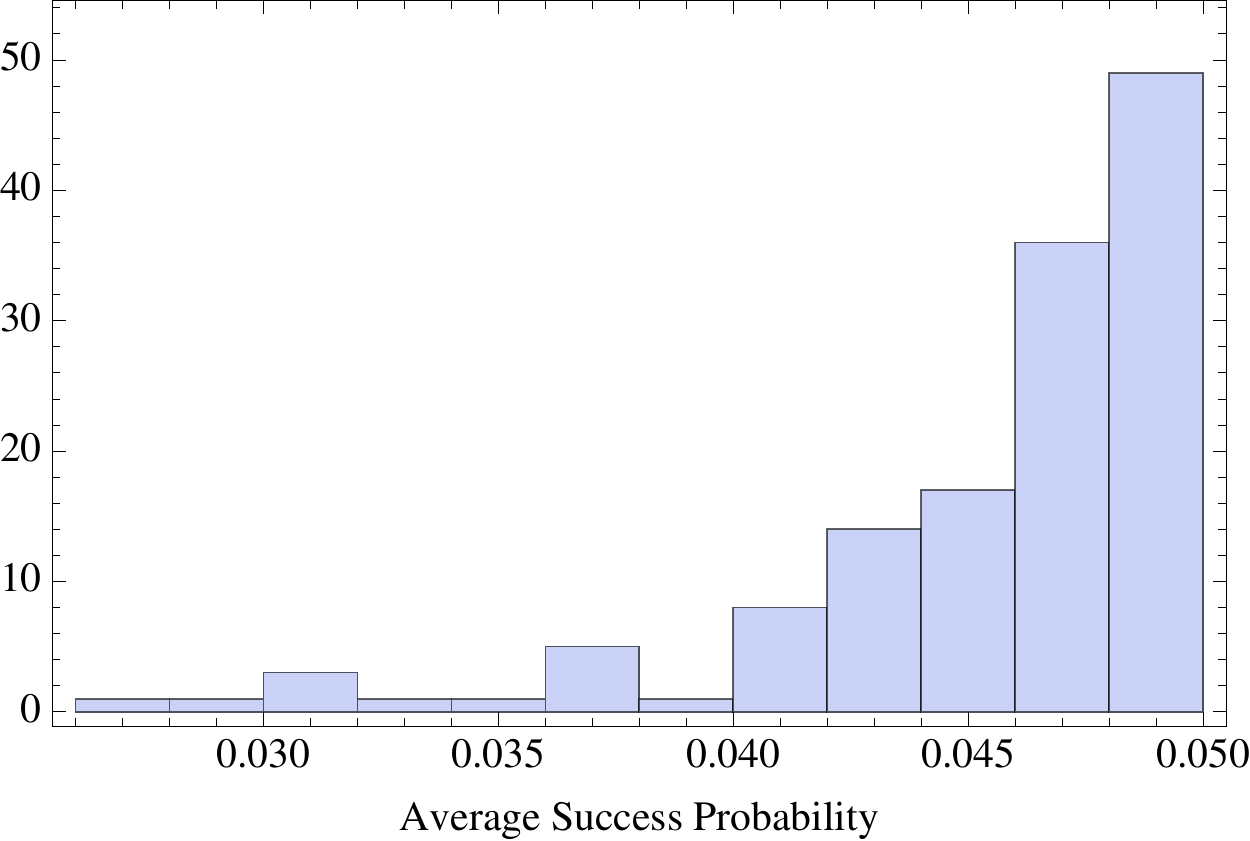}
\par\end{centering}
\caption{The average success probability at $T = 100$ for 137 instances obtained by initializing the system in each of the 20 first excited states of $H_B$. \label{fig:excitedSuccesses}}
\end{figure}

\begin{figure}[H]
\begin{centering}
\includegraphics[scale=1,trim=0 0 0 0]{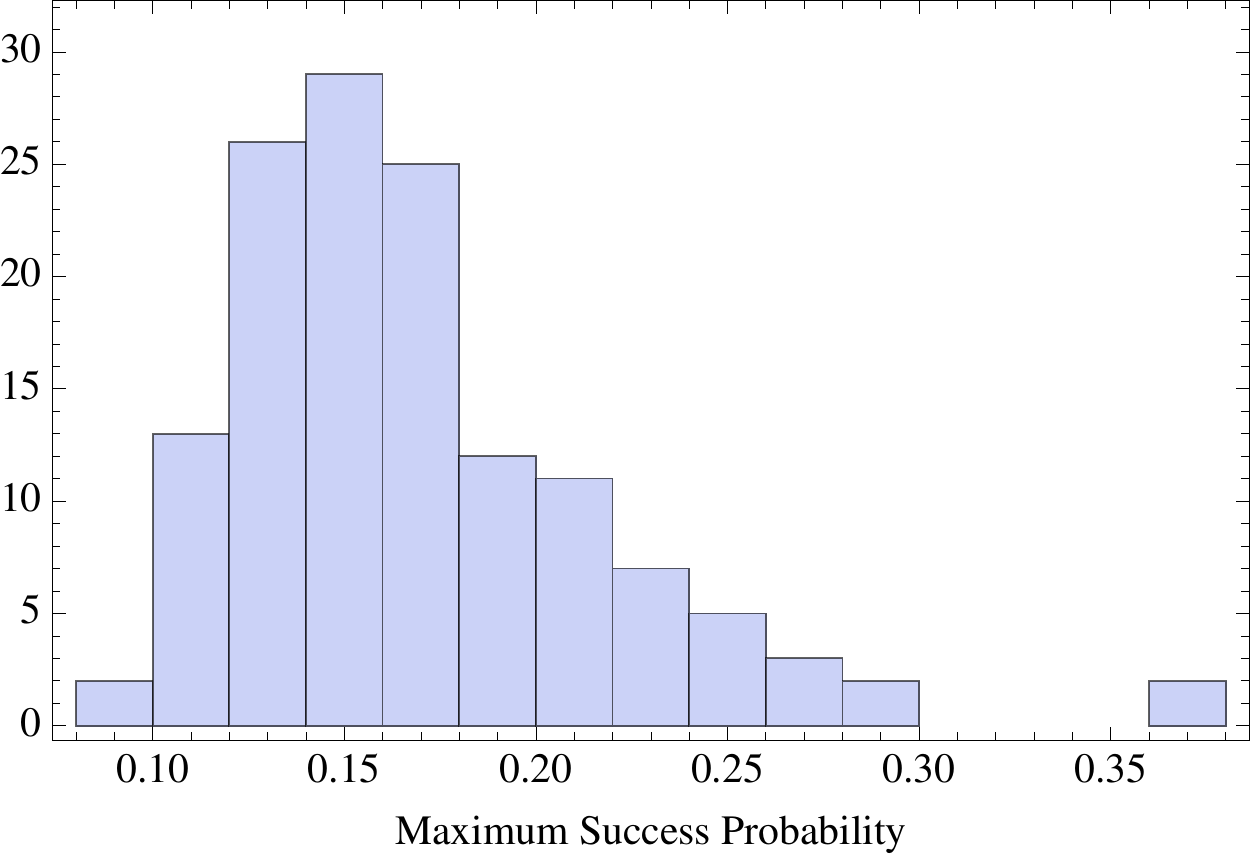}
\par\end{centering}
\caption{The maximum success probability at $T = 100$ for 137 instances obtained by initializing the system in each of the 20 first excited states of $H_B$.\label{fig:excitedSuccessesMax}}
\end{figure}

In the previous section we saw that having a substantial overlap with the first excited state before the avoided crossing increases the overlap with the ground state at the end of the evolution.  In this section we attempt to directly exploit this effect by preparing the system at $t = 0$ to be in one of the 20 first excited states of $H_B$, obtained by taking the ground state (in equation \ref{psi:init}) and flipping one of its qubits from $(|0\rangle+|1\rangle)/\sqrt{2}$ to $(|0\rangle-|1\rangle)/\sqrt{2}$.   We did this for each of the 20 first excited states for each of our $137$ hard instances.  For each instance the average success probability over the 20 excited states of $H_B$ is given in figure \ref{fig:excitedSuccesses}, and the maximum success probability for every instance is given in figure \ref{fig:excitedSuccessesMax}.  
 
 As shown in figure \ref{fig:excitedSuccesses}, this strategy produces an average success probability near $1/20$ for most of our 137 instances. This saturates the upper bound given by probability conservation, since the sum of the success probabilities associated with the 20 orthonormal initial states cannot exceed 1.  

A similar strategy to ours was used in \cite{nagaj-2012} to overcome an exponentially small gap in a particular Hamiltonian construction by initializing the system in a random low energy state.  The authors argue that this technique is useful whenever there are a small number of low lying excited states that are separated from the remaining space by a large energy gap.  The possibility of using non-adiabatic effects to drive a system from its ground state on one side of a phase transition to its ground state on the other side was considered in \cite{caneva-2011} as a problem in quantum control theory.   Here we quantify the viability of this strategy for particularly hard instances of MAX 2-SAT at 20 bits.  

\section{The Meandering Path May Be Faster}
The traditional time-dependent Hamiltonian in equation \ref{eq:H} represents a path in Hamiltonian space which is a straight line between $H_B$ and $H_P$.  Here, as was previously considered in \cite{farhi-2002}, we modify this path by adding an extra randomly chosen Hamiltonian $H_E$,
\begin{equation}
H(t) = \left(1-\frac{t}{T}\right)H_B + \frac{t}{T}\left(1-\frac{t}{T}\right)H_E + \frac{t}{T} H_P.
\end{equation}
A reasonable constraint on $H_E$ is that it be a sum of local terms with the same interaction graph as the problem Hamiltonian $H_P$, but should not use any other information specific to the particular instance.   We consider three categories of $H_E$:
 %In the case of MAX 2-SAT this means that $H_E$ will have 1 and 2-local terms for every pair of bits which are involved in a clause.
\begin{enumerate}
\item {\bf Stoquastic} with zeroes on the diagonal.  Each 2-local term of $H_E$ is a linear combination of 1 and 2-qubit Pauli operators from the set $\{I\pa{x}, \pa{x}I, \pa{z}\pa{x}, \pa{x} \pa{z}, \pa{x}\pa{x}, \pa{y}\pa{y}\}$.  For each 2-local term, the 6 real coefficients are sampled from a Gaussian distribution with mean zero, and are then normalized so that their squares sum to 1.  Moreover, the coefficients are kept only if the local Hamiltonian term constructed in this way is stoquastic (i.e. all of the off-diagonal matrix elements are real and non-positive).  
\item {\bf Complex} with zeroes on the diagonal.  Each 2-local term of $H_E$ is a linear combination of 1 and 2-qubit Pauli operators chosen from the set $\{I\pa{x}, \pa{x}I, I\pa{y}, \pa{y}I, \pa{z}\pa{x} , \pa{x} \pa{z}, \pa{x} \pa{x}, \pa{y}\pa{y}, \pa{z}\pa{y}, \pa{y}\pa{z}, \pa{y}\pa{x}, \pa{x}\pa{y} \}$.  For each 2-local term, the 12 real coefficients are sampled from a Gaussian distribution with mean zero, and are then normalized so that their squares sum to 1.
\item {\bf Diagonal}.   Each 2-local term of $H_E$ is a linear combination of 1 and 2-qubit Pauli operators chosen from the set $\{I\pa{z}, \pa{z}I, \pa{z}\pa{z}\}$.  For each 2-local term, the 3 real coefficients are sampled from a Gaussian distribution with mean zero, and are normalized so that their squares sum to 1.  
\end{enumerate}

The reason that we work with zero diagonal $H_E$ in the first two categories is to be sure that we are exploring purely quantum strategies for increasing the success probability, since the diagonal elements of $H_E$ could be seen as time-dependent classical modifications to the energy landscape of $H_P$.   The reason that we separate stoquastic path change from general complex path change is that ground states of stoquastic Hamiltonians have various special properties which may limit their computational power.    Ground state local Hamiltonian problems are known to have lower computational complexity when the Hamiltonians are restricted to be stoquastic \cite{bravyi-2006}\cite{bravyi-2006-2}.  Moreover, ground state properties of stoquastic Hamiltonians can be determined using Quantum Monte Carlo (a collection of classical methods for finding properties of quantum systems) at system sizes of up to a few hundred qubits (for a general review see \cite{sandvik-2011}, for an application to the QAA see \cite{farhi-2012}).   Non-stoquastic Hamiltonians have a ``sign problem'' that prevents this, and we know of no efficient simulation techniques for non-stoquastic Hamiltonians at system sizes of more than roughly 20 qubits.  The traditional QAA Hamiltonian defined by equations \ref{eq:Hp}, \ref{eq:HB}, and \ref{eq:H} is stoquastic, and we are interested in seeing whether non-stoquastic path change can increase the computational power of this algorithm.  

\begin{figure}[H]
\begin{centering}
\includegraphics[scale=.9,trim=20 15 0 5]{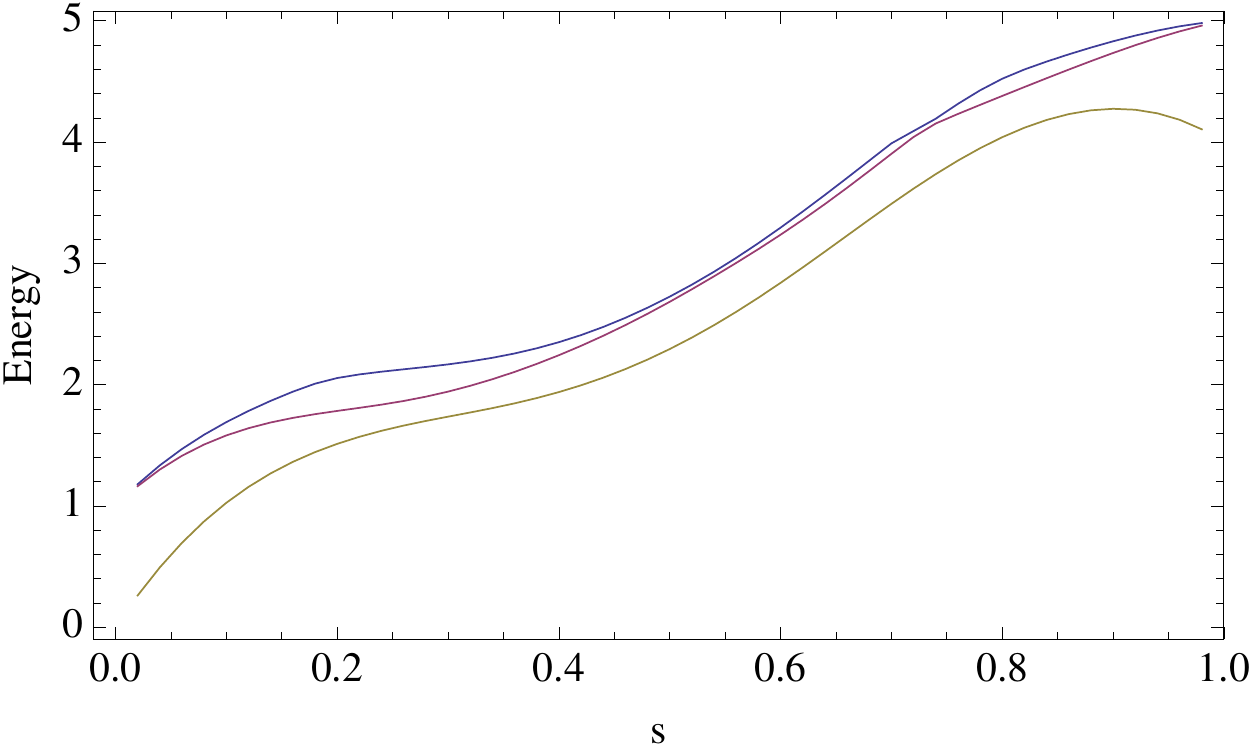}
\par\end{centering}
\caption{The energy spectrum of instance $\#1$ with a particular choice of complex $H_E$ which gives $P(100) = 0.91$.\label{fig:spectrumComplex}}
\end{figure}

As a first demonstration of the potential for path change to increase success probabilities, we return to instance $\#1$ which had $P(100) = 5\times 10^{-5}$.  In figure \ref{fig:spectrumComplex} we plot the spectrum for this instance with a particularly successful choice of complex $H_E$, and see that the avoided crossings in figure \ref{fig:spectrum137} have been eliminated.

\begin{figure}[H]
\begin{centering}
\includegraphics[scale=1.2,trim=10 80 0 0]{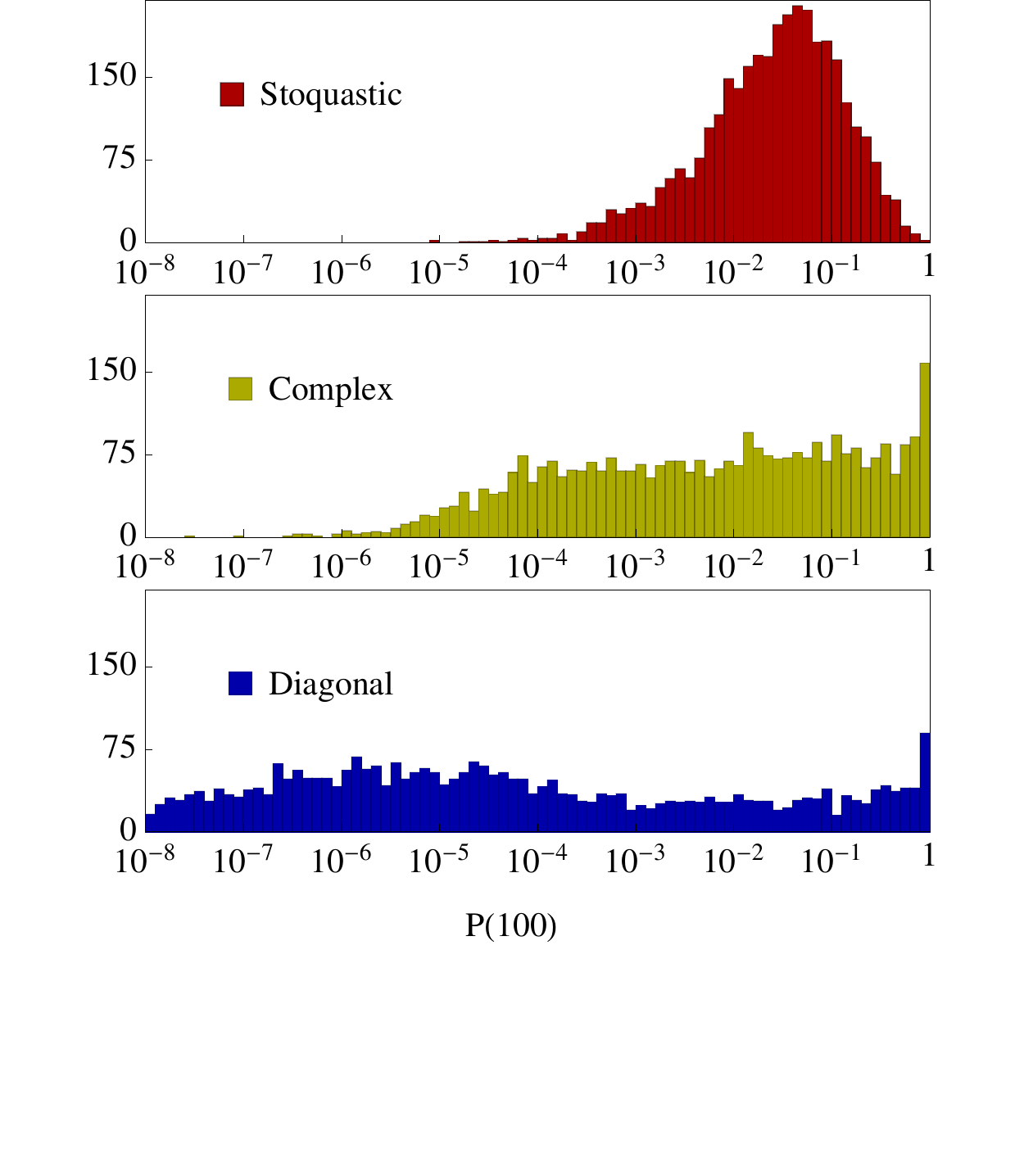}
\par\end{centering}
\caption{The distribution of success probabilities for 137 hard instances, when each is run with 25 randomly sampled path changes.\label{fig:pathChangeAll}}
\end{figure}

We tested the performance of this strategy by simulating $25$ trials of stoquastic, complex, and diagonal path change for each of our $137$ hard instances (which all have $P(100) < 10^{-4}$ when $H_E = 0$).  The path changes are all chosen independently so that there are no correlations between the instances.   Simulating these path change trials for all of our instances required over 25,000 hours of CPU time.  The full distribution of success probabilities we obtained at $T = 100$ is given in figure \ref{fig:pathChangeAll}. 

\begin{figure}[H]
\begin{centering}
\includegraphics[scale=1.2,trim=20 80 0 0]{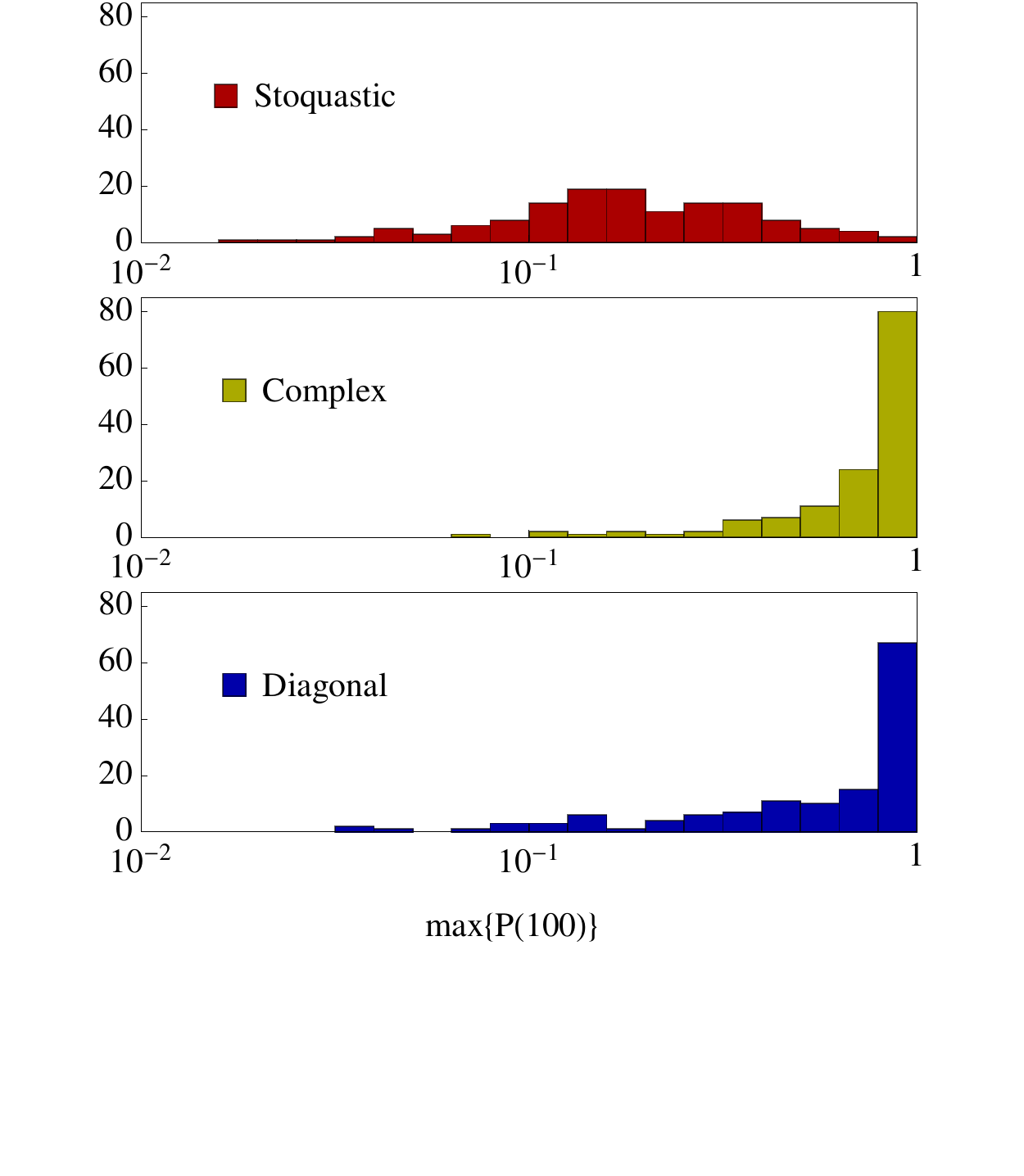}
\par\end{centering}
\caption{The maximum success probabilities for each of the 137 hard instances, when each is run with 25 randomly sampled path changes.\label{fig:pathChangeMax}}
\end{figure}

While stoquastic path changes almost always increase the success probability above $10^{-4}$, we see that they rarely produce success probabilities near $1$.   This is shown in the distribution of the maximum success probability we obtained for each instance with 25 trials of path change, shown in figure \ref{fig:pathChangeMax}.  

\begin{figure}[H]
\begin{centering}
\includegraphics[scale=1.2,trim=20 80 0 0]{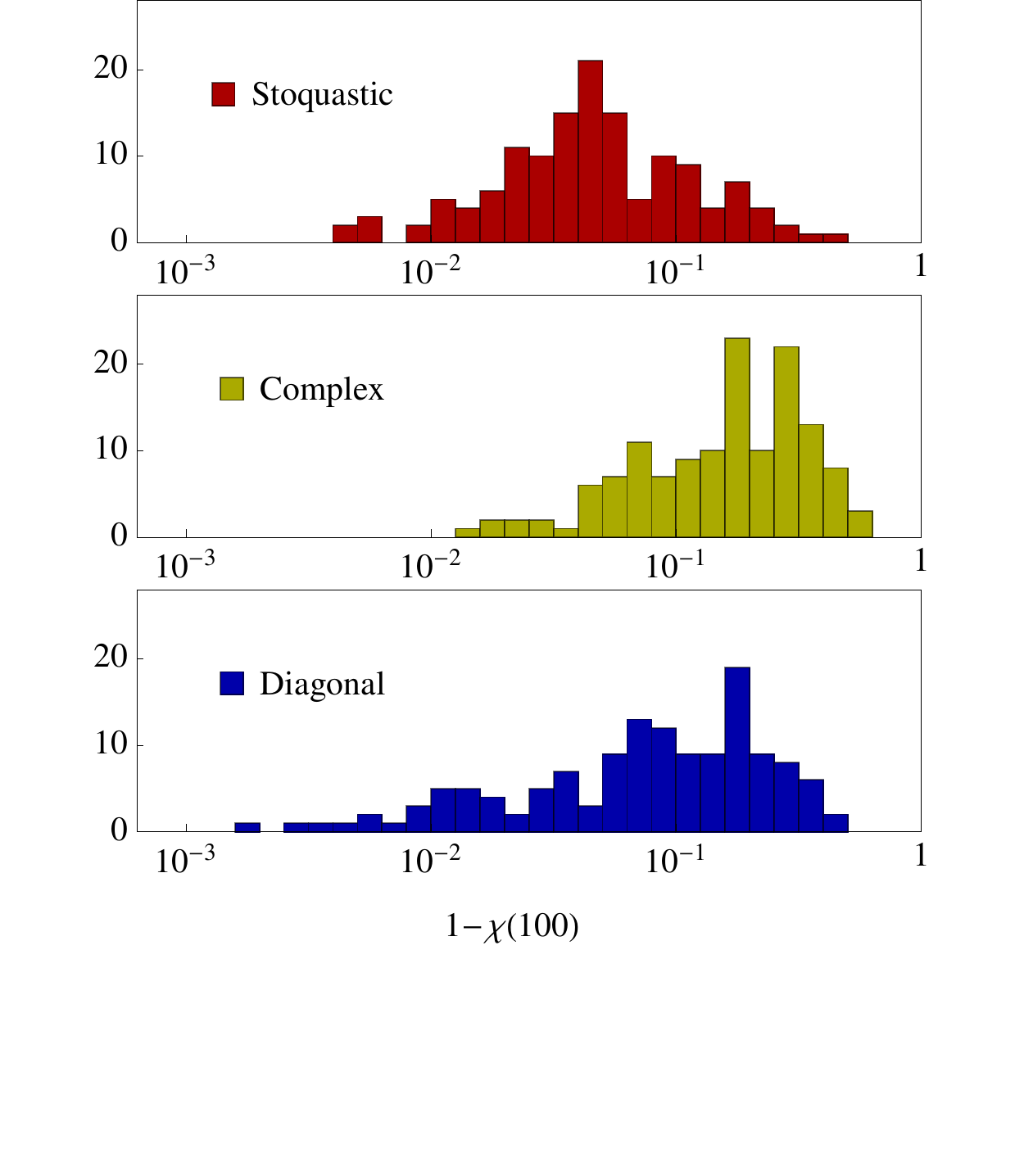}
\par\end{centering}
\caption{The effective success probabilities (given by $1$ minus the geometric mean of the failure probabilities) obtained by running each of the 137 hard instances with 25 randomly sampled path changes.\label{fig:pathChangeChi}}
\end{figure}

To take into account the spread in the distribution we estimate the geometric mean of the failure probabilities obtained by many trials of path change.  For each instance we use the 25 trials to compute
\begin{equation}
\chi = \left ( \prod_{i=1}^{25} \text{failure probability of the }i\text{-th trial} \right)^{1/25}.
\end{equation}
We take $1 - \chi$ to be the effective success probability of a single trial of the path change strategy.  In figure \ref{fig:pathChangeChi} we give the distributions of $1 - \chi$ for our ensemble of 137 hard instances.  

We find that all three types of path change increase the effective success probability for all 137 of our hard instances, with complex path change typically producing the largest increase in the effective success probability.  

\begin{figure}[H]
\begin{centering}
\includegraphics[scale=.85,trim=25 370 0 20]{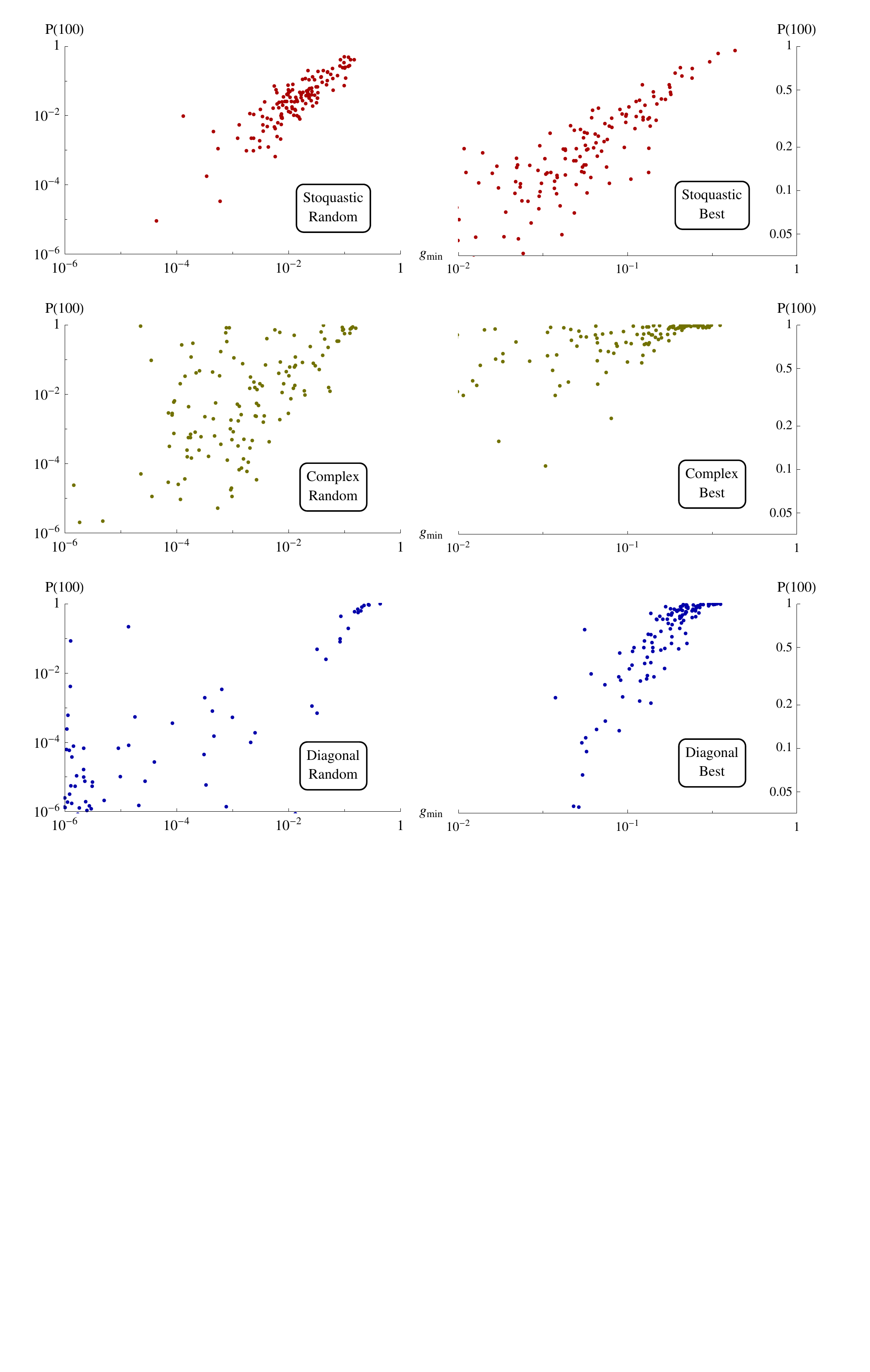}.
\par\end{centering}
\caption{A comparison of success probabilities with the minimum spectral gap for several trials of path change.   The plots in the left column contain one random path change trial for each instance.  In the right column we plot the most successful path change trial for each instance.  Note that the scales for the probabilities and the minimum gaps are different between the left column (random) and the right column (best).  “ \label{fig:gvs}}
\end{figure}

To check whether the widening of the spectral gap seen in figure \ref{fig:spectrumComplex} also occurs for other successful trials of path change, we computed the minimum spectral gap $g_{min}$ between the ground state and the first excited state for a subset of our path change trials.  For each of our 137 hard instances we computed $g_{min}$ for the most successful path change trial of each of the three types, and also for a randomly selected trial of each of the three types.  Figure \ref{fig:gvs}  compares these minimum gaps to the corresponding success probabilities.

We see a correlation between high success probability and large gaps, and almost no data with large gaps and low success probabilities.  

\section{Conclusion}

We generated over 200,000 instances of MAX 2-SAT on 20 bits with a unique optimal satisfying assignment, and selected the subset for which the numerically exact time-simulation of QAA governed by equations \ref{eq:Hp} through \ref{eq:Schrodinger} finds success probabilities of less than $10^{-4}$ at $T=100$.  We gave three strategies which increase the success probability for all of these instances.  First we ran the adiabatic algorithm more rapidly, and observed an increased success probability at shorter times for all 137 instances. Second, we initialized the system in a random first excited state of $H_B$ and saw that the average success probability for this strategy is close to the upper bound $0.05$ for the majority of hard instances.  Finally, we observe that adding a random local Hamiltonian to the middle of the adiabatic path often increased the success probability, and that different types of path changes produced different distributions, with the stoquastic case most often increasing the success probability, the complex case being the most likely to give success probabilities close to $1$, and the diagonal case having the most spread and highest likelihood of reducing the success probability. 

To guard against the possibility that what we observe are low bit number phenomena, we also tested these strategies for the QAA version of the Grover search algorithm.  Here the problem Hamiltonian $H_P$ assigns energy $1$ to all of the computational basis states aside from one of them which is assigned energy $0$.  The Grover problem requires exponential time for any quantum algorithm, so we expect that our strategies should not improve the success probability.  Indeed, at $n = 20$ qubits they do not.  

One striking thing about these strategies is that they increase the success probability for all of the very low success probability instances we generated.   This may be a consequence of testing our strategies on particularly hard instances, which have the most room for improvement. Figure \ref{fig:successDistribution} shows that the overwhelming majority of instances we generated at 20 bits are far easier than the ones we selected. At higher bit number it may be that most instances have very low success probability when the traditional QAA is run for a time that scales polynomially in the number of bits.  If the only effect of these strategies is to bring the most difficult instances in line with the typical instances, which may in fact have very low success probabilities, then the algorithmic value of these strategies is limited.  We hope that one day these strategies will be tested on a quantum computer running the Quantum Adiabatic Algorithm at high bit number where classical simulations are not available.  

\section{Acknowledgements}
We are very grateful to Christoph Paus for allowing us to use the CMS Tier 2 distributed computing cluster, and to William Detmold for allowing us to use his Lattice QCD cluster.  We appreciate the assistance with using these resources that we received from Maxim Goncharov and Andrew Pochinsky.  We thank Aram Harrow and Sam Gutmann for useful discussions.  EF would like to thank the members of the Quantum Artificial Intelligence Lab at Google for helpful suggestions.  This work was supported by the US Army Research Laboratory's Army Research Office through grant number W911NF-12-1-0486, the National Science Foundation through grant number CCF-121-8176, and by the NSF through the STC for Science of Information under grant number CCF0-939370.  CYL gratefully acknowledges support from the Natural Sciences and Engineering Research Council of Canada.  EC was funded by NSF grant number CCF-1111382 and did this work while a visiting student at the MIT CTP.

\bibliographystyle{plain} \nocite{*} \bibliographystyle{plain}
\bibliography{ImprovingQAO}

\end{document}